\documentclass[aps,prl,twocolumn,groupedaddress, showpacs]{revtex4}
\usepackage{graphicx}
\usepackage{amssymb}
\usepackage[squaren, cdot]{SIunits}

\begin{document}


\title{Reconfigurable, site-selective manipulation of atomic quantum systems in two-dimensional arrays of dipole traps}


\author{J. Kruse, C. Gierl, M. Schlosser, and G. Birkl}
\email[]{gerhard.birkl@physik.tu-darmstadt.de}
\affiliation{Institut f\"{u}r Angewandte Physik, Technische Universit\"{a}t Darmstadt, Schlossgartenstra\ss e 7, 64289 Darmstadt}


\date{\today}
\newcommand{\bra}[1]{\ensuremath{\langle #1 |}}
\newcommand{\ket}[1]{\ensuremath{| #1 \rangle}}
\newcommand{\braket}[2]{\ensuremath{\langle #1 | #2 \rangle}}

\newcommand{\beq}{\begin{equation}}
\newcommand{\eeq}{\end{equation}}

\newcommand{\state}[3]{\ensuremath{^{#1}}\textrm{#2}\ensuremath{_{#3}}}
\newcommand{\atom}[2]{\ensuremath{^{#1}}\textrm{#2}}

\newcommand{\sfrac}[2]{\nicefrac{\ensuremath{#1}}{\ensuremath{#2}}}
\newcommand{\om}{\omega}
\newcommand{\pii}{2\pi\cdot}
\newcommand{\lk}{\left(}
\newcommand{\rk}{\right)}
\newcommand{\tn}{\textnormal}
\newcommand{\tg}[1]{\ensuremath{_{\textnormal{#1}}}}

\begin{abstract}
We trap atoms in versatile two-dimensional (2D) arrays of optical potentials, 
prepare flexible 2D spin configurations, perform site-selective coherent manipulation, and demonstrate the implementation of simultaneous measurements of different system properties, such as dephasing and decoherence. This novel approach for the flexible manipulation of 
atomic quantum systems is based on the combination of 2D arrays of microlenses and 
2D arrays of liquid crystal light modulators.
It offers novel types of control for the investigation of quantum degenerate gases, quantum information processing, and quantum simulations. 

\end{abstract}
%

\maketitle



\begin{figure}[tbp]
\includegraphics[width=0.8\linewidth]{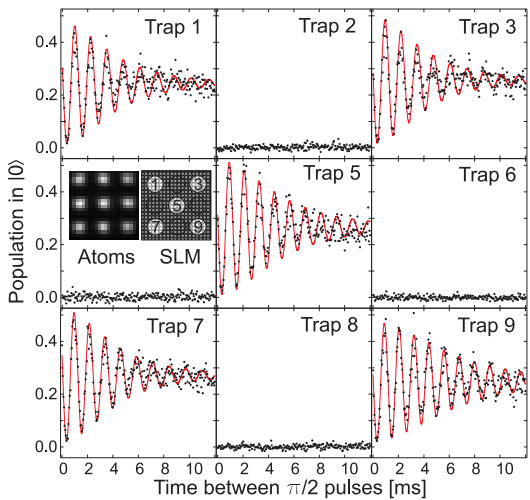}
 \caption{Ramsey oscillations in a site-selectively addressed two-dimensional dipole trap array. The panels show the population in state \ket{0} in nine traps (all loaded with atoms - see left inset) as a function of the free evolution time. The light used for inducing the Ramsey oscillations is controlled by a spatial light modulator and is applied to the traps with odd order number exclusively (right inset). Only in the addressed traps, Ramsey oscillations are observed. No crosstalk to not addressed traps is visible. Each data point is averaged 5 times.\label{Schach}}
 \end{figure}

\begin{figure}[tbp]
\centering
\includegraphics[width=0.9\linewidth]{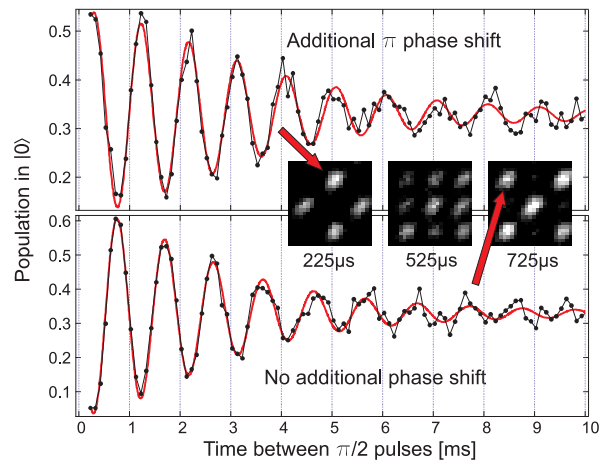}
\caption{Ramsey oscillations with a phase difference of $\pi$ at two neighboring sites 
as expected in a 2D configuration of anti-parallel spins.
Inset: atoms detected in state  \ket{0}. The two interleaved subsets of atom samples prepared in a 2D anti-parallel spin configuration, oscillate with an expected phase difference of $\pi$.
Each data point is averaged 5 times.
\label{Bilder_Schach.eps}}
\end{figure}

\begin{figure}[tbp]
\includegraphics[width=0.8\linewidth]{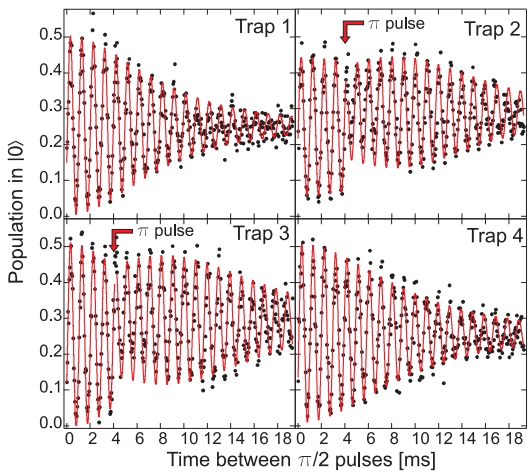}
 \caption{Site-selective coherent manipulation (Ramsey and spin-echo method) in a 2D array of dipole traps. The oscillations in four neighboring traps are shown. During the coherent evolution, an additional $\pi$ pulse is applied to traps No. 2 and No. 3 after \unit{4}{ms}, switching the phase by $\pi$ without influencing traps No. 1 and No. 3. Rephasing of the signal in traps No. 2 and No. 3 due to a spin-echo is clearly visible. Each data point is averaged 5 times. 
\label{Echo_LCD}}
\end{figure}

Optical dipole potentials 
such as optical lattices or arrays of focused laser beams
provide flexible geometries for the synchronous investigation of multiple atomic quantum systems, as studied e.g in the fields of quantum degenerate gases or quantum information processing with atoms \cite{Bloch_optical_lattices, Anderlini_NIST_lattice, Meschede, dumke:097903, Bergamini_Grangier:2004:JOptSocAmB, yavuz_Saffman:2006:PRL}.
In comparison, optical lattices provide a larger number of potential wells (up to $10^6$) \cite{Bloch_optical_lattices, Anderlini_NIST_lattice, Meschede},
but the required ability of performing flexible site-selective addressing is still a challenge \cite{karski:053001, Weiss_2007_Nature, bakr-2009_Greiner, Ott_scanning_electron_microscopy_Nature2008}.
On the other hand, architectures based on two-dimensional arrays of tightly focused laser beams  inherently provide the ability to address single sites \cite{Birkl2001, dumke:097903, Bergamini_Grangier:2004:JOptSocAmB} at the expense of a smaller number of wells (up to several $10^4$) and a larger separation of sites (typically several $\mu m$). 
Significant future progress is expected from complementing the advantages of these configurations, namely the scalability and the ability to perform quantum operations in parallel, with an additional versatility by achieving reconfigurable, site-selective initialization, manipulation, and detection of individual quantum systems at each site.
 \begin{figure}[bp]
 \includegraphics[width=0.9\linewidth]{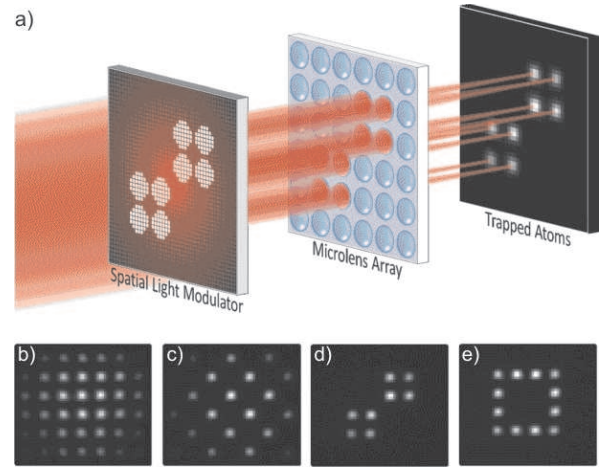}
 \caption{(color online) (a) A liquid crystal display is used as addressible spatial light modulator for illuminating reconfigurable sections of an array of microlenses. This produces versatile two-dimensional arrays of dipole traps.
(bottom) Fluorescence images of small samples of trapped atoms. (b) Fundamental trap configuration with all microlenses illuminated and all traps filled with atoms. (c) - (e) Reconfigured trap patterns with selectively illuminated microlenses for creating (c) 'superlattices', (d) trap structures for quantum error correction or plaquette states, and (e) a ring lattice with periodic boundary conditions.
Images are averaged 20 times.   
\label{render}}
 \end{figure}
\\
In this work, we introduce and experimentally implement a novel approach towards  this goal: we trap and coherently manipulate two-dimensional (2D) sets of atomic quantum systems in flexible and reconfigurable architectures. We combine 2D arrays of microlenses with per-pixel addressable spatial light modulators (SLM) (Fig. \ref{render}). This results in reconfigurable, per-site addressable 2D arrays of diffraction-limited laser foci in the focal plane of the microlens array. 
By re-imaging we reduce the structure (separation \unit{55}{\mu m}, spot size \unit{3.7}{\mu m}) while maintaining diffraction-limited performance.
Although our current setup is limited to a minimum structure size of about \unit{1.3}{\mu m} by its numerical aperture (NA), with optics of sufficiently high NA \cite{karski:053001, Weiss_2007_Nature, bakr-2009_Greiner}, 2D arrays of laser foci with sub-micron stucture size can be achieved. 
Thus, our approach allows to create two-dimensional arrays of optical micro-potentials for combined trapping and addressing purposes but also for matching sub-micron-period optical lattices with a flexible system for 2D site-selective addressing. 
\\
On the next pages, we present versatile trap configurations produced in a robust fashion and demonstrate the ability to allocate atoms in flexible sets of dipole traps with each trap controlled separately (Fig. \ref{render}).
In addition, for the first time experimental results on performing site-selective, but also simultaneous coherent manipulation of a two-dimensional set of atomic quantum systems are presented. 
This allows us to initialize, manipulate, and readout the quantum state in each trap individually, in subsets, or globally. 
Central to our approach is the fact that we use the SLM only for addressing individual microlenses, but not as a holographic phase element for creating complex focal spot structures \cite{Bergamini_Grangier:2004:JOptSocAmB, boyer_Foot:2006:PRA}.
This ensures high stability and a diffraction-limited light field in the focal plane, both given by the advantageous characteristics of the microlenses.
For the same reason, we do not use a dynamically reconfigured SLM for the transport of atomic quantum systems (see \cite{boyer_Foot:2006:PRA}), but rather have implemented atom transport in an independent fashion in our previous work using beam scanning techniques \cite{lengwenus-2009}.
\\
We demonstrate the key properties of this approach in our experiment on quantum information processing (QIP) \cite{Nielsen_chuang} with two-dimensional arrays of atomic quantum bits (qubits) \cite{dumke:097903, Birkl_LPhys}.
For the work in this paper, we use small atom samples (10 to 100 atoms per site) with a separation of \unit{55}{\mu m} as qubits, although we have achieved freely selectable trap separations down to 0 \cite{lengwenus-2009} and
single atom preparation in two-dimensional trap arrays as well \cite{Schlosser}. 
In our previous work \cite{dumke:097903, Birkl_LPhys, lengwenus-2009, Schlosser}, we have demonstrated most of the key features necessary for the successful implementation of QIP in our architecture with only the realization of a two-qubit gate remaining to be shown. 
There is a clear path for achieving this by e.g using the long-range interactions between well separated Rydberg atoms which was successfully implemented in pairs of dipole traps recently \cite{PhysRevLett_Saffman_CNOT_RydbergGate, PhysRevLett_grangier_RydbergEntanglement}. The necessary trap separation in the range of \unit{5-10}{\mu m} can easily be achieved in our architecture as shown in \cite{lengwenus-2009}.
\\
A schematic view of our setup is presented in Fig. \ref{render} (a). Laser light for atom trapping or manipulation globally illuminates a two-dimensional SLM which is placed in front of an array of $50\times 50$ microfabricated
refractive lenses of which a subset of about 50 lenses is used typically.
The microlenses have a diameter of $\unit{100}{\mu m}$, a
pitch of $\unit{125}{\mu m}$, and a focal length of $\unit{1}{mm}$. Microlenses with a wide range of specification are available from various sources. 
The SLM allows for the separate control of the light power impinging on each microlens by inscribing a reconfigurable pattern of transmitting or non-transmitting disks on a dark background into the SLM. The disks are imaged on individual microlenses and the illuminated lenses produce a 2D array of diffraction-limited spots in the focal plane. 
The focal plane is re-imaged into a glass cell based vacuum system using a telescope
consisting of an achromatic lens ($f=\unit{80}{mm}$) and a diffraction limited
lens system ($f=\unit{35.5}{mm}$, NA$=0.29$). This results in a spot pattern with a pitch of $\unit{55}{\mu m}$ and a measured waist of $w_0=\unit{(3.7\pm 0.1)}{\mu m}$ (1/e$^2$ radius) consistent with the NA used. Fully exploiting our maximum available NA of 0.29, a waist below \unit{1.3}{\mu m} could be reached.
\\ 
Inside the vacuum cell, rubidium ($^{85}$Rb) atoms are trapped and cooled in a standard magneto-optical trap (MOT). During
a sequence of optical molasses the atoms are transferred into the superimposed 2D
array of laser foci which act as a 2D array of dipole traps for light red-detuned from the D1 and D2
transitions of Rb. 
As trapping laser we use a titanium-sapphire laser at a wavelength of \unit{795.8}{nm} for the experiments presented in Fig. \ref{render} and \unit{815}{nm} for the ones presented in Figs. \ref{Schach} - \ref{Echo_LCD}. 
  Atom detection is achieved by resonant fluorescence imaging
using the MOT beams for illumination and collecting
the fluorescence light with an intensified CCD camera.
\\
We use a liquid crystal display (LCD) taken from a
commercial data projector as spatial light modulator (SLM). 
The LCD is a 2D array of pixels, each acting as an
individually tunable retardation waveplate. 
We use the LCD followed by a polarizing beam splitter as a per-pixel intensity modulator. Our device is a $1024\times$\unit{768}{pixel} array with a total active area of $20\times \unit{15}{mm^2}$ ($19\times \unit{19}{\mu m^2}$ per pixel) operated in transmission. We measured a rise time of $\unit{60}{ms}$ and a fall time of $\unit{10}{ms}$. For faster switching times, SLMs based on ferroelectric liquid crystals or micro-mechanical mirrors can be used. The light pattern transmitting the SLM is imaged with a demagnification of a factor of 2 onto the microlens array. An area of 80 pixels corresponds to the area of one single microlens. Due to the small pixel size, light scattering into several diffraction orders occurs. To maintain the maximum spatial resolution, only the lowest diffraction order is utilized, higher orders are blocked by an iris. The contrast between maximum and minimum transmission can be optimized by a  $\lambda /2$ waveplate between the LCD and the polarizing beam splitter. For optimized contrast of 270:1 we measure a total transmission efficiency of \unit{5.9}{\%} including all losses. It is possible to increase the transmission efficiency at the expense of reduced contrast. Due to the ability of controlling the transmission in each pixel in 256 steps by the VGA output of a standard computer, we can control the relative transmitted intensity 
in the range between \unit{0.4}{\%} and \unit{100}{\%}. 
\\
We use this setup to produce versatile two-dimensional configurations of atom traps. 
In Fig. \ref{render} (bottom) fluorescence images of small samples of atoms trapped in various configurations are shown. 
Atoms are only trapped in those dipole traps which correspond to the microlenses illuminated through the SLM.  
Figure \ref{render} (b) shows the fundamental structure of the 2D trap register, created by globally illuminating the microlens array (all pixels of the SLM turned to full transmission). 
With the SLM, we have the ability to change the pitch and orientation of the grid of the dipole trap array by illuminating only every other microlens (Fig. \ref{render} (c)) creating a 'superlattice' with definable structure. Another possibility is to generate subsets of smaller, separated dipole trap arrays (Fig. \ref{render} (d)), which allow to realize schemes for quantum error correction \cite{Nielsen_chuang} or plaquette states in two-dimensional lattice spin models \cite{Zoller_PlaquetteStates}. Finally, Fig. \ref{render} (e) shows atoms trapped in a 2D configuration comparable to a ring lattice with periodic boundary conditions \cite{PhysRevLett.95.063201_Ringlattice,olmos:043419_PRA2009_Rydberg}. 
As can be seen, this scheme of producing arbitrary trap patterns is very flexible on one hand and very stable and robust on the other hand: due to the fact that we are always using the microlens array to define the underlying structure in the focal plane, the stability and diffraction-limited performance of the focal structure are not compromised by the added flexibility through the SLM.
\\ 
In addition to creating flexible trap geometries, we also perform coherent manipulation of 2D sets of atomic quantum systems in parallel  as well as site-selectively in a reconfigurable fashion. This capability is essential for scalable approaches towards QIP.
In our work, qubit states are represented by hyperfine substates of the $5 S_{1/2}$ ground state of $^{85}$Rb. To be insensitive to fluctuations of magnetic fields to first order, we use the two clock states $\lk\ket{0}=\ket{F=2,\ m_F=0},\ \ket{1}=\ket{F=3,\ m_F=0}\rk$ and coherently couple them using the light of two phase-locked diode lasers. 
%
The two lasers are about \unit{20}{GHz} red-detuned with respect to the D2 line at \unit{780}{nm}. The pulse length of both beams is controlled by an acousto-optical switch. A typical duration of an applied $\pi$-pulse is $\unit{200}{\mu s}$.  State-selective detection is performed by removing the atoms in $\ket{F=3}$ by a laser pulse which is resonant to the $\ket{F=3}\rightarrow \ket{F\,'=4}$ transition and subsequently detecting the remaining $\ket{F=2}$ atoms. For site-selectivity, we send the coupling laser beams onto the atoms by illuminating a second microlens array through an SLM. Both arrays have identical specifications. Their focal planes are transferred into the vacuum cell after superimposing them with a dichroic mirror. For inscribing freely configurable phase shifts into each trap,
we can adjust the control beam intensity through each SLM pixel separately
using the 256 steps in transmission.
\\
In the experiments on coherent manipulation (Figs. \ref{Schach} - \ref{Echo_LCD}) 
the gaussian shaped laser beam illuminating the microlens array used for trapping 
has a power of $\unit{(137\pm 2)}{mW}$ and a 1/e$^2$-radius of $\unit{(700\pm 4)}{\mu m}$. This yields a power of $\unit{(1.23\pm 0.04)}{mW}$ and a trap depth of $k_B\times \unit{(60\pm 2)}{\mu K}$ in the central trap.
In Fig. \ref{Schach} we show a $3\times 3$ section of simultaneous Ramsey experiments \cite{lengwenus-2009} with the coherent coupling light field configured in a checkerboard pattern (see Fig. \ref{render} (c)).  
We observe Ramsey oscillations only in the addressed traps without detecting any measurable oscillations in the ones not addressed, although there are atoms in all traps (see inset of Fig. \ref{Schach}). 
The Ramsey oscillations show the well known reduction of contrast with time due to inhomogeneous dephasing \cite{Kuhr_Meschede:2005:PRA}.
We do not observe any measurable cross talk between neighboring sites. 
Based on the measured intensity contrast of the SLM (1:270), we infer that a light field giving a $\pi$ rotation in the addressed traps leads to a $4.2 \cdot 10^{-3} \cdot \pi$ rotation in the not addressed traps.
\\
The site-selective addressability also allows for the preparation of complex two-dimensional spin configurations. Such systems are of extreme interest for studying complex quantum states and their interactions, such as antiferromagnetic ordering \cite{Hofstetter_AntiferromagneticOrdering} or multipartite entanglement with atom-light interfaces \cite{stasinska-2009_Sanpera}.
Here, we use the SLM to prepare a 2D configuration of periodically changing anti-parallel spins by applying a $\pi$ phase shift in the pattern of Fig. \ref{render} (c) to atoms initially in state \ket{1} at all sites. To demonstrate the coherent site-selective reversal of spins, a Ramsey experiment is performed in all traps simultaneously after the site-selective spin-flip operation. In Fig. \ref{Bilder_Schach.eps} a sequence of three fluorescence images showing atoms in state \ket{0} after different free evolution times is presented for nine traps (inset) and Ramsey oscillations in two neighboring traps (specified by arrows) are given in detail. All traps show Ramsey oscillations, but due to their different starting spin states, we observe the expected phase difference of $\pi$ in the Ramsey oscillations between addressed and not addressed sites. 
\\
Finally, we prove the ability to coherently manipulate quantum superposition states in a site-selective fashion by performing interleaved Ramsey and spin-echo experiments in a checkerboard configuration.
The spin-echo method is an extension of the Ramsey method
with an additional $\pi$ pulse between the $\pi/2$ pulses. In Fig. \ref{Echo_LCD} we use global coupling laser beams for applying the two $\pi/2$ pulses to all atom samples in a 2D register simultaneously. The additional $\pi$ pulse, addressing every other site, is applied after $T_\pi=\unit{4}{ms}$ via an independent pair of coupling laser beams controlled by the SLM. As expected, we observe a $\pi$ phase shift and the change from a Ramsey to a spin-echo signal with its typical rephasing behavior at \unit{8}{ms} at the addressed sites.
Since the decay of the Ramsey signal with increasing free evolution time gives information on dephasing whereas the decay of the spin-echo signal with increasing $T_\pi$ gives information on decoherence, we have implemented a method to gain information on both of these important properties simultaneously. 
\\
In conclusion, we have presented a versatile, scalable and reconfigurable architecture for neutral atom trapping and quantum state manipulation.
It is based on site-selectively addressable registers of focused laser beams which are created by combining arrays of microlenses with two-dimensional spatial light modulators. 
In this fashion, we add the flexibility of the SLM to the stability and the diffraction-limited performance of the microlens array. We have implemented atom trapping in reconfigurable 2D trap patterns and the simultaneous as well as site-selective coherent qubit manipulation. Combined with our abilities of reducing the separation of sites down to the single micron level, of single atom detection \cite{Schlosser} and of coherent quantum state transport \cite{lengwenus-2009}, this approach lends itself to the further development of successful architectures for quantum information processing, quantum simulation, and the investigation of quantum degenerate gases.
\\ 
This work was supported financially in part by the Deutsche Forschungsgemeinschaft (DFG), by the Deutscher Akademischer Austausch Dienst (DAAD), by the European Commission (Integrated Project SCALA), and by IARPA and NIST (Award 60NANB5D120).

\end{document}